\documentclass[12pt]{article}
\usepackage{amsmath,amssymb,graphicx}
\usepackage{amsthm}
\usepackage{eufrak}

\newtheorem{prop}{Proposition}

\newcommand{\pr}{\noindent{\bf Proof}. }
\newcommand{\rem}{\noindent{\bf Remark}. }
\newcommand{\rems}{\noindent{\bf Remarks}. }

\newcommand{\pa}{\partial}

\newcommand{\const}{\textrm{const}}

\newcommand{\hs}{ \hspace{1cm}}

\newcommand{\im}{\textrm{Im}  }
\newcommand{\re}{\textrm{Re} } 
\newcommand{\B}{\Big}
\newcommand{\blan}{\Big  \langle} 
\newcommand{\bran}{\Big  \rangle} 
\newcommand{\sq}{\square}
\newcommand{\inn}{\textrm{in}} 
\newcommand{\out}{\textrm{out}}

\newcommand{\be}{\begin{equation}}
\newcommand{\ee}{\end{equation}}

\newcommand{\al}{\alpha}
\newcommand{\De}{\Delta}
\newcommand{\de}{\delta}

\newcommand{\Om}{\Omega}
\newcommand{\om}{\omega}

\newcommand{\ep}{\epsilon}

\newcommand{\cC}{{\cal C}}
\newcommand{\cD}{{\cal D}}

\newcommand{\cO}{{\cal O}}

\newcommand{\cH}{{\cal H}}
\newcommand{\cS}{{\cal S}}

\newcommand{\cF}{{\cal F}}

\newcommand{\bbR}{ \mathbb{R} }

\begin{document}
 
\title{Quantum radiation from a classical point source}
\author{ J. Dimock \\ Department of Mathematics\\ SUNY at Buffalo} 
\maketitle

\begin{abstract}
We study the radiation of photons from a classical charged particle.      We particularly  consider a situation where the particle has
a constant velocity in the distant  past, then is accelerated, and then has a constant velocity in the distant future.   Starting with 
no photons in the distant past  we seek to characterize the quantum  state of the photon field  in the distant future.  Working in the Coulomb gauge and in a C* algebra formulation,   we give sharp conditions 
on whether this   state is or is not in Fock space. 
\end{abstract} 

%\tableofcontents

\section{Introduction} 

We study the radiation of the  quantum electromagnetic field  in the presence of a classical  source.  
With a specified four-current consisting of functions   $ ( j^0, j)   $   satisfying  $\pa_{\mu} j^{\mu} =0$,  one  seeks  associated  quantum field operators     $A=( A^0, A) $.   In the Lorentz gauge $\pa_{\mu} A^{\mu}  =0$  these are connected by the  wave equation with source
$\square A^{\mu} = j^{\mu}$.      A     textbook treatment  of this problem  can be found in Itzykson-Zuber \cite{ItZu80} whose sign conventions we follow.  A mathematical treatment is due
to   Naudts-DeRoeck  \cite{NaDe04}  .    A characteristic phenomena is that    
if one starts in a state with no photons in the distant past, the  vacuum in Fock space,   it may  evolve into a state in the distant future  with
infinitely many photons radiated  and   not   in Fock space.    This is  sometimes known as the infrared catastrophe. 

In this paper we are interested in characterizing exactly when this happens.  We particularly study the case of a point source
where the four-current has the form 
 \begin{equation} 
 \label{point} 
 \begin{split} 
 j^0 (x,t)  = &   \de (x-x(t))    \\
 j(x,t)   = &   x' (t) \de ( x - x(t) )  \\
 \end{split}
 \end{equation}
and $x(t)$ is the trajectory of the charged particle in $\bbR^3$.   We suppose that
the particle has a velocity  $v_{in}$ in the distant past,   then accelerates for a finite amount of time,  and then continues with a velocity $v_{out}$ in the distant future.  We start   with the Fock vacuum in the distant past considered as a state $\om_{in} $  on 
the $C^*$-algebra generated by the fields.  It evolves to
 state  $\om_{out} $ in the distant future.    We establish the following results
 \begin{itemize}
 \item   If  $v_{out} = v_{in} =0$ and the trajectory $x(t)$ is smooth then $\om_{out} $ is a Fock state.
 \item  If  $v_{out} = v_{in} =0$ and the velocity $x'(t)$  has a discontinuity  then $\om_{out} $ is not  a Fock state.
\item   In general for a smooth trajectory $\om_{out} $ is a Fock state if and only if $v_{in} = v_{out} $.
 \end{itemize}

 We work
in the Coulomb gauge rather than the Lorentz gauge. The advantage is that we avoid working with indefinite metric Hilbert spaces.   
The disadvantage is that  relativistic properties like Lorentz invariance and locality are not manifest, although they are still there.  (See
Skagerstam, Erikson, Rekdal \cite {SER19a} for a discussion of locality in the Coulomb gauge.)

The  Coulomb gauge is defined by the condition $ \nabla \cdot  A =0$ on the magnetic potential.   
In this gauge Maxwells equations become
 \begin{equation} \label{Maxwell} 
 \begin{split}
 -\De A^0  =&  j^0 \\
\frac{\pa^2 A}{\pa t^2}  - \De A   + \nabla \frac{\pa}{\pa t}   A^0 =&  j  \\
\end{split} 
\end{equation}
The first equation for the electrostatic potential $A^0$  has the solution 
\be \label{two} 
A^0(x,t)  = \int  V(x-y)  j^0  (t, y) d y 
\ee
where $V(x) = (4\pi)^{-1} |x|^{-1} $ is the Coulomb potential. 
In the second equation we call the term   $ \nabla ( \pa A^0/\pa t ) $ the longitudinal current.    Since  $ \pa j^0 /\pa t  +   \nabla  \cdot  j       =0$  
this can be expressed as 
\be
 j_L(x,t)   =    \nabla ( \pa A^0/\pa t )(x,t)   =  -  \nabla   \int V(x-y)  ( \nabla \cdot j ) (y,t) dy  
\ee
It satisfies $\nabla \cdot j_L = \nabla \cdot j$. 
The transverse current is defined by  $ j_T  =  j - j_L  $ and it satisfies $\nabla \cdot j_T  =0$.
The second equation in (\ref{Maxwell})  and the constraint  can now be written 
\be \label{Maxwell2} 
 \frac{\pa^2 A}{\pa t^2}  - \De A    =  j_T  \hs    \nabla \cdot  A =0
\ee
or just  $\sq A = j_T$ with $\sq = \pa_t^2 -\De$.

The electrostatic potential $A^0$ is given by (\ref{two}) and is not quantized.   We seek quantum field operators  $A$ 
satisfying (\ref{Maxwell2}).

\section{Preliminaries}

\subsection{fundamental solutions}
 We review some facts about solutions of the wave equation with source.   At first this is just classical so 
  we   seek  functions    $A $ satisfying $ \sq  A  = j$.   We are interested in solutions of the form   $A = G_{\pm}* j$ where $G_{\pm}$ are retarded and  advanced   fundamental solutions  which we define shortly.    We want to allow $j$ to be certain  distributions so    $G_{\pm}* j$ is defined as 
  $<G_{\pm}* j, \phi> = <j, G_{\mp} * \phi> $ for $\phi$ in (vector-valued) $\cS(\bbR^4)$, the Schwartz space of smooth rapidly decreasing functions. 
      However  we also want to work at sharp time
  so the test functions $\phi$ will be replaced  by $f \otimes \de_t$ with $f$  in (vector-valued) $\cS(\bbR^3)$.   The upshot is 
  that we want to define   functions   $   G_{\pm} *  (f \otimes \de_t) $, or without the convolution    $G_{\pm}  (f \otimes \de_t)\equiv   G_{\pm } (f,t  )  $.

The $G_{\pm} $ are given by  
\be
G_{\pm} (x,t) = - \frac{1} { (2\pi)^4} \int_{p_0 \in \bbR \pm  i \ep}  e^{ i(- p_0 t + px)}  \frac{1}{p_0^2- |p|^2}   dp_0 dp
\ee    
For $f \in  \cS(\bbR^3) $ this is to be understood as
for $ G_{\pm}(f,t)  =    \int  f(x) G_{\pm}(x,t) dx$ given by 
\be 
G_{\pm}(f,t)  =-  (2 \pi)^{-\frac52}  \int_{p_0 \in \bbR \pm i \ep}   \frac{1}{p_0^2- |p|^2} e^{-ip_0t} \tilde f(-p)  \  dp_0 dp
\ee
where $\tilde f(p) = (2\pi)^{-\frac32} \int e^{-ipx} f(x) dx$  is the Fourier transform of $f$. 

 Consider   $G_{+}(f,t) $.   We have  $|e^{-i p_0t}| = e^{\im( p_0)  t}$ so 
if    $t<0$  we can   close the $p_0$-contour in the upper half plane and get zero.   On the other hand  If    $t>0$  we    we close the contour in the lower half plane, picking up 
poles at $p_0 = \pm \om $   where $\om(p) = |p|$. Similarly for $G_{-}(f,t) $.    We obtain 
\be
 G_{\pm  } (f,t  ) 
  = \pm (2 \pi )^{-\frac32}    \theta(\pm t)  i   \int \B(  e^{-i \om t}   - e^{i \om t}   \B) \tilde f(-p)   \frac{dp}{ 2 \om} 
 \ee
 where $\theta$ is the Heaviside function.

 The convolution is defined by 
 $
 ( G_{\pm } *  (f \otimes \de_{t_0}) )(x,t ) = G_{\pm } \B(  f(x-  \cdot ),\de_{t_0} (t- \cdot)\B)
 $. 
 But the  Fourier transform of $f(x- \cdot  )$ at $-p$ is   $e^{ipx} \tilde f(p)$ and $\de_{t_0} (t-\cdot )= \de_{t-t_0}  $.
Thus we have  
   \be \label{sister}  
( G_{\pm  } *  (f \otimes \de_{t_0}) )(x,t )
    =  \pm (2 \pi )^{-\frac32}       \theta( \pm(t-t_0)) i  \int \B(  e^{-i \om (t-t_0) }   - e^{i \om (t-t_0 ) }   \B)e^{ipx}   \tilde f (p)    \frac{dp}{ 2 \om}        
\ee    
 
 We  also will want to consider the propagator  $G=G_+-G_-$.   In this expression the Heaviside functions $\theta$ disappear 
 and we have
 \be \label{sharp} 
( G *  (f \otimes \de_{t_0}) )(x,t ) =  u_{f} (x,t-t_0)
\ee
where
\be \label{sharp2} 
   u_{f} (x,t)  = (2 \pi )^{-\frac32}   i   \int  \B(  e^{-i \om t }   - e^{i \om t }   \B)e^{ipx}   \tilde f(p)     \frac{dp}{ 2 \om} \ee    
is  a smooth solution of the wave equation with  data at $t= 0 $
\be \label{sinkhole}
u_{f}(x,0) = 0
  \hs 
 \frac{\pa u_{f}}{ \pa t}(x,0)= f(x) 
\ee

 \subsection{free quantum field}

    We  review  the mathematics of the   quantized free field in the Coulomb gauge.
In this case we can take  $A^0 =0$, and    study the field equation
\be
\frac{\pa^2 A}{\pa t^2}  - \De A =0   
\ee
We  want to define a quantum field operator satisfying  the equation,
the constraint $\nabla \cdot  A =0$,  and the canonical commutation relations.

It is convenient to use the  relativistic  Hilbert space   which is  (vector valued)
\be  
\cH = L^2(\bbR^3,(2 \om)^{-1}  dp )
\ee 
again with $\om(p) =|p|$.  On  this space     define 
a projection operator. 
\begin{equation}
(Ph)_j (p) =     \sum^3_{k=1} \B(\de_{jk} - \frac{ p_jp_k}{|p|^2} \B) h_k(p) 
\end{equation}
 which  is   the orthogonal  projection onto  the space of   transverse functions 
satisfying $p\cdot h(p) =0$.   
The  one photon Hilbert space   is $\cH_T = P \cH$,
 the n-photon Hilbert space    $\cH_T^n$ is the n-fold symmetric tensor product of $\cH_T$ with itself, and the Fock space over $\cH_T$ is  
\be 
\cF( \cH_T)  =\bigoplus_{n=0}^{\infty} \cH_T^n
\ee

For $h \in \cH_T$ let  
  $a^*(h) $  and  $a(h)$    be the  creation and annihilation operators 
on the Fock space,  with $a(h)$ anti-linear in $h$ and
$a^*(h) $ linear in $h$.  These operators  (and all operators in this section) are defined on the dense  subspace $\cD$ consisting  
elements of  Fock space with a finite number of particles.  They satisfy  $[a(h_1),a^*(h_2 )] = (h_1,h_2)$.  We also define momentum space fields  $A_{\circ}(h) $ and $\Pi_{\circ}  (h) = A_{\circ} (ih)$
by 
\be \label{twist} 
\begin{split}
A_{\circ}( h)    = &  \frac{1}{\sqrt{2}}  \B( a^*(h) + a(h)  \B ) \\
 \Pi_{\circ}( h )    = &  \frac{i}{\sqrt{2}}  \B( a^*(h) - a(h)  \B ) \\
\end{split} 
\ee
These  are real linear in $h$,  symmetric,  and satisfy    $[A_{\circ}(h_1), A_{\circ}(h_2) ]  = i \im  ( h_1,h_2) $.

The field operator is defined as a distribution.   For  a  real test function $f \in \cS (\bbR^3)$
  define  $A(f,t) $  (formally  $\int f(x) A(x,t) dx$) by  
\begin{equation}  \label{ingot}
 A( f, t)    =  \sqrt{2}   A_{\circ}  \B( e^{i \om t}   \tilde f_T \B ) =    a^*( e^{i \om t}    \tilde f_T  )  + a( e^{i \om t}   \tilde f_T  )  
\end{equation} 
where $\tilde f_T  \equiv P \tilde f$.   This satisfies the field equation 
 \be
\frac{ \pa^2 A}{ \pa t^2} (f,t) - A( \De f, t) =0
 \ee
   It also satisfies  $\nabla \cdot A =0$ since  for  scalar $g \in \cS(\bbR^3) $ we have     $(\nabla \cdot A)(g,t)  \equiv -  A( \nabla  g,t)  $  and
    $  (\nabla g)_T ^{\sim }(p) = i P(p) p   \tilde g (p)    =0$. 
We have   the commutation relations 
  \begin{equation}\label{key}
    [ A( f_1, t), A( f_2, t)] =  2 \im (\tilde   f_{1,T},  \tilde f_{2,T} )=0 
  \end{equation}
  The vanishing  follows since for $f$ real $ \overline{ \tilde f (p) } = \tilde f (-p) $ so $ (\tilde   f_{1,T},  \tilde f_{2,T} )$ is real.

 The field has a conjugate momentum  $  \Pi ( h, t) = \pa/ \pa t  \  A( h, t) $ given by 
 \begin{equation}\label{keyII}
  \Pi ( f, t)   = \sqrt{2}  \Pi_{\circ} \B(   e^{i \om t}   \om  \tilde f_T \B )  =   i \B(  a^*( e^{i \om t}  \om   \tilde f_T  )  -  a( e^{i \om t}  \om   \tilde f_T  )  \B) 
   \end{equation}
The $ \Pi ( f, t)  $  commute with each other  and satisfy 
 the canonical commutation relations  (CCR) 
  \begin{equation}\label{keyIII}
   [  A( f_1, t)  ,\Pi( f_2,t ) ] = i (\tilde f_{1,T}, 2\om \tilde f_{2,T})_{\cH}  = i  (\tilde f_{1,T},   \tilde f_{2,T})_{L^2(\bbR^3, dp) }      
  \end{equation}

\subsection{the $C^*$ algebra and coherent states} 
 
 We want to allow states for this system which are not Fock states, and for this  we focus on the algebra of the fields.
See for example  Bratteli-Robinson \cite{BrRo81} for more details on  this standard material.  We consider operators $A_{\circ}(h)$ for $h$ in a dense subspace
 $ \cS \subset \cH_T$.    The operator   is  essentially self-adjoint on  the domain $\cD \subset \cF(\cH_T) $, and so 
 we can  define
\begin{equation}\label{key2}
 W_{\circ}(h) = e^{ i  A_{\circ} (h )  }  \hs \hs   h \in  \cS 
\end{equation}  
 This satisfies  
\begin{equation}\label{key3}
 W_{\circ}(h_1) W_{\circ}(h_2) = e^{ -\frac{i}{2}  \im (h_1,h_2)  } W_{\circ}(h_1 + h _2)
\end{equation}  
 This is the Weyl form of the CCR.
The operators $W_{\circ} (h ) $ generate a $C^*$ algebra, denoted $ \EuFrak{A} $,  which is the closure of the set of operators  which are finite sums  $\sum_n c_nW_{\circ}(h_n ) $ 
 in the Banach space of bounded operators on $\cF(\cH_T) $.   Different choices of the dense subspace $\cS$ give different algebras,  and 
 we will make a specific choice later. 
 
  A state $\om$  on the algebra is a positive linear functional of norm one.  Positivity is the condition that   $\om( B^*B) \geq 0$ for any $B \in \EuFrak{A} $.      Any  normalized state in Fock space defines a state on the algebra.
 In particular the   Fock vacuum $\Om_0$   defines a state by $\om_0(\  \cdot \ )  = ( \Om_0, [\ \cdot \ ] \Om_0) $. 
 It satisfies  $a(h) \Om_0 = 0$  and   is characterized by
 \be \label{monty} 
 \om_0 (W_{\circ}(h) )  =(\Om_0,  e^{i A_0(h) } \Om_0)  = ( \Om_0, e^{i  a^*(h)/\sqrt{2}  }  e^{i  a(h)/\sqrt{2}  } \Om_0)   e^{-\frac14 \| h \|^2}  
 = e^{-\frac14 \| h \|^2}  
 \ee

Another class of states are the coherent states.   Given $J \in \cH_T$  and the Fock vacuum $\Om_0$ we define the coherent state
\be \label{coherent}  
\Om_J    =     e^{-\frac12 \|  J \|^2}  e^{ a^*(J)  } \Om_0  
=   e^{-\frac12 \| J  \|^2}  \sum_{n=0}^{\infty} \frac{1}{n!}   (  a^* (J ) )^n\Om_0   
\ee
The series converges in Fock space  since $\|  (  a^* (J ) )^n\Om_0   \| = \sqrt{n!} \| J \|^n$.  It is normalized to $\| \Om_J  \| =1$
and  is an eigenstate of the annihilation operator: 
\be
a(h) \Om_J  =  (h,J)  \Om_J   
\ee
As a state on the algebra   $\om_h = (\Om_h ,  [\ \cdot \ ] \Om_h) $
it satisfies 
 \be \label{monty2} 
 \begin{split} 
 \om_{J}  (W_{\circ} (h) )  = & (\Om_J ,  e^{i A_0(h) } \Om_J )\\
   = &  ( \Om_J, e^{i  a^*(h)/\sqrt{2}  }  e^{i  a(h)/\sqrt{2}  } \Om_J )   e^{-\frac14 \| h \|^2}  \\
     = & e^{ i \overline{(h,J)} /\sqrt{2} }   e^{ i (h,J)/\sqrt{2} }      e^{-\frac14 \| h \|^2}  \\
   =&  e^{-\frac14 \| h \|^2   + i\sqrt{2} \re (J, h) }  \\
\end{split} 
 \ee

More generally we define a coherent state on $\EuFrak{A} $  to be a state satisfying
\be
\om ( W_{\circ} (h) )   =  e^{-\frac14 \| h \|^2   + i\sqrt{2} \re L(h) }  
\ee
for some linear  function $L$ on the dense domain $\cS$.   It can also be written 
\be
\om ( W_{\circ} (h) )   =  \om_0 ( W_{\circ}(h) )  e^{i \sqrt{2} \re L(h) }  
\ee
The  positivity condition is satisfied since  for any finite sequence of complex numbers $c_n$  and $h_n \in \cS$ and $B = \sum_n c_n W_{\circ}(h_n)$
we have 
\be
\begin{split} 
\om( B^*B)  =  &\sum_{n,m}  \bar c_m c_n \  \om \B( W_{\circ} (h_n - h_m) \B)  e^{ \frac{i}{2}  \im (h_n,h_m) } \\
=   &\sum_{n,m} \B( \bar   c_m e^{- i  \sqrt{2}\re L( f_m) }  \B)  \B( c_n e^{i  \sqrt{2}\re L( f_n) }   \B)  \ \om_0 \B( W_{\circ} (h_n - h_m) \B)  e^{ \frac{i}{2}  \im (h_n,h_m) }  \\
= & \om_0(B_0^* B_0) \geq 0 \\
 \end{split} 
 \ee
 where
 \be   
 B_0 =  \sum_n c_n e^{i  \sqrt{2}\re L( f_n) }  W_{\circ} (h_n) 
 \ee
 
  The  coherent state  $\om$ defined by $L$  is  known to be  a pure state (Honegger-Rapp \cite{HoRa90}).  
Furthermore   $\om$ is a Fock
 state if and only if $L$ is continuous on the pre-Hilbert space $\cS$ (Honegger-Rieckers \cite{HoRi90}).  This is the criterion we will use  in the following.

 \newpage

 \section{Regular   source} \label{regular}

 To warm up we first consider our problem with a regular source. 
 We take $j  \in \cC^{\infty}_0 (\bbR^4) $ and    seek  quantum field operators which solve
$\sq A  =   j_T $  and $  \nabla \cdot A =0 $.
   Note  that  partial Fourier transform   $ \tilde j(p,t) $ is    bounded and rapidly decreasing in $p$  as is 
  $ \tilde j_T(p,t) = P(p)\tilde  j(p,t)  $ 
  The inverse Fourier transform $j_T( x,t) $ then  has compact support in $t$,  and for each $t$ is bounded and continuous in $x$ (since  $ \tilde j(p,t) $
  is integrable in $p$) and square integrable in $x$ (since  $ \tilde j(p,t) $ is square integrable in $p$).   The localization of   $j_T( x,t) $  in $x$ is 
  considerably weaker
  than that of $j(x,t)$.

We want  a solution which is a free field  in the distant past   and the distant future.    Therefore 
 we take $A_{\inn}  (f, t ) $ to be a free field  defined   from creation and annihilation
 operators $a^*_{\inn} (h), a_{\inn} (h) $ as in  (\ref{ingot}). 
 Then   for real $f \in \cS(\bbR^3) $ we define 
 \be    \label{zucchini}
 A (f,t)   = A_{\inn} (f,t)  +  (  G_+  * j_T)( f,t) 
 \ee
   The expression $G_+ *  j _T$ can be analyzed pointwise.  Indeed we have explicitly
    \be \label{omnibus}  
( G_{\pm}  *  j_T) (x,t )
    =    \pm (2\pi)^{-\frac32}  i \int     \theta( \pm(t-s)) i   \B(  e^{-i\om( t-s) }   - e^{i \om(t-s)  t}    \B) e^{ipx} \tilde j_T( p,s)    \frac{dp}{ 2 \om}     \ ds   
\ee 
Then    $( G_+  * j_T)(x,t)  $   vanishes  in the distant past   since  $j_T(p,s)$ does,  and hence
 $A (f,t)   = A_{\inn}(f, t) $ in the distant past. 
We have     $\sq  ( G* j_T )  = j_T$, also in the sense of distributions,  and hence $ \sq A  = j_T$.   Furthermore    $\nabla \cdot     ( G* j_T ) $, also in the sense of distributions,  and hence        $\nabla \cdot A = 0$

  Next   we define $ A_{\out }  (f,t)  $ by     
 \be \label{squash} 
 A (f,t)   = A_{\out } (f,t)  +   (G_- *  j _T)(f,t) 
  \ee
  Then $ A(f,t)     = A_{\out }(f,t)   $ in the distant future since $(G_- *  j _T)(f,t) $ vanishes in the distant future. 
 Combining (\ref{zucchini}) and (\ref{squash})  and identifying  $G= G_+ - G_-$ gives 
 \be \label{bingo1}
  A_{\out } (f,t)  =   A_{\inn }  (f,t)  + (G   *  j_T) (f,t)
 \ee
 where now
  \be
  \begin{split} 
   \label{omnibus2}  
( G *  j_T)(x,t )
 &   =  (2\pi)^{-\frac32}  i  \int  \B(  e^{-i\om (t-s)  }   - e^{i \om (t-s) }    \B) e^{ipx} \tilde j_T( p,s)    \frac{dp}{ 2 \om}     \ ds   \\
 &  = (2 \pi)^{-1}  \  i  \int  \B(  e^{-i\om t  } \hat j_T(\om, p )    - e^{i \om t }  \hat j_T(-\om, p)    \B) e^{ipx}     \frac{dp}{ 2 \om}   \\
\end{split} 
\ee    
  where  $\hat j$ is the four dimensional Fourier transform with Lorentz inner product:
  \be
  \hat j(p_0,  p  ) = (2\pi)^{-2} \int e^{ i( p_0t - px) } j(x,t)  \  dx dt = (2 \pi)^{-\frac12} \int e^{ip_0t} \tilde j (p,t) dt
  \ee
Then  $G *  j_T$ satisfies   $ \sq  ( G *  j_T) =0 $   and $\nabla \cdot  (G *  j_T)  =0$, and hence so does      $A_{\out } (f,t) $  
  Furthermore 
 together with 
 $\Pi_{\out} (f,t)  =  \pa A_{\out} (f,t) / \pa t$
it satisfies the CCR.  Thus $A_{\out } (f,t) $ is a free field.

In the second term in (\ref{omnibus2}) make the change of variable $p \to - p$.  Since  $j$ is real we have 
  $  \overline{\hat j_T(\om, p)  } = \hat j_T(-\om, - p) $  and
the second term is  identified as the complex conjugate of the first term.   
Thus if we define
\be
J(p) =  - \sqrt{2 \pi} \  i  \   \hat j(\om, p )  
\ee
and $J_T(p) = P( p) J(p) $
then we have 
\be
( G *  j_T)(x,t )
 =- (2\pi)^{-\frac32}  2   \ \re \    \int     e^{-i(\om t- px)   } J_T(p)   \frac{dp}{ 2 \om} 
\ee  
 This gives
 \be
( G *  j_T)(f,t )
 =  -2  \ \re \  \int   \tilde   f(-p) e^{-i\om t } J_T(p)   \frac{dp}{ 2 \om}   =  - 2  \ \re \ (e^{i \om t}  \tilde  f_T, J_T) 
\ee  
Now we have 
\be
 A_{\out} (f,t) = A_{\inn} ( f,t) -  2  \ \re \ (e^{i \om t}  \tilde  f_T, J_T) 
\ee
 
 We define creation and annihilation operators  by taking what amounts to  the positive and negative frequency parts  of the last expression.
 We  define for $h \in \cH_T$
   \be  \label{xmas1}
   \begin{split}  
 a^*_{\out} (h)  = &  a^*_{\inn} (h) - (J_T,h)   \\  
 a_{\out} (h)  =&   a_{\inn} (h)  -  (h, J_T)   \\ 
 \end{split} 
 \ee
 These satisfy the standard  $ [a_{\out} (h_2)  ,  a^*_{\out} (h_2) ]= (h_1,h_2) $.  
Then 
 \be   \label{xmas2}
   \begin{split}  
 a^*_{\out} (e^{i \om t} \tilde f_T)  = &  a^*_{\inn} (e^{i \om t} \tilde f_T) - (J_T,e^{i \om t} \tilde f_T)   \\  
 a_{\out} (e^{i \om t} \tilde f_T)  =&   a_{\inn} (e^{i \om t} \tilde f_T)  - (e^{i \om t} \tilde f_T,J_T)   \\ 
 \end{split} 
 \ee
 If we add these we get  $A_{in} ( f,t) -  2  \ \re \ (e^{i \om t}  \tilde  f_T, J_T)  = A_{out} (f,t) $ on the right.   Thus we must get the 
 same  on the left which says 
 \be  \label{xmas3}
  A_{\out} (f,t)  = a^*_{\out} (e^{i \om t} \tilde f_T)+ a_{\out} (e^{i \om t} \tilde f_T)
\ee 
 just as for   $  A_{\inn} (f,t)  $   .

The out vacuum in Fock space  should satisfy    $a_{\out}(h) \Om_{\out} = 0$.  
 Note that  $j \in \cC^{ \infty}_0 (\bbR^4) $ implies that $\hat j$  is bounded and rapidly decreasing, hence  $J_T \in  \cH_T$.   
 Therefore we can construct   the coherent state  
\be \label{coherent2}  
\Om_{\out}     = \Om_{J_T}  =      e^{-\frac12 \|  J_T \|^2}  \exp ( a_{ in} ^*(J_T)  ) \Om_{\inn}  
\ee
This gives the   the desired
\be
a_{\out}(h) \Om_{\out} = \B( a_{\inn} (h) - (h,J_T)  \B)\Om_{\out}  =0 
\ee
We summarize:

\begin{prop} 
Let $j  \in \cC^{\infty}_0(\bbR^4)$.  Then $ J_T(p)  = -  \sqrt{2\pi} \  i \  \hat j_T(  \om ,p)$ is in $\cH_T$ and  the out 
vacuum $ \Om_{out}$ exists in Fock space. 
\end{prop}  
\bigskip

\rems
\begin{enumerate}
\item 
   To obtain this result  can certainly relax the condition that $j \in \cC^{\infty}_0 (\bbR^3)$.  The key condition is   is  $J_T  \in \cH_T  $  which follows
   from $J \in \cH$  which says
\be
\int  |\hat j (   \om , p ) |^2  \frac{dp}{2 \om}  < \infty
\ee
This is similar to a condition that  Naudts - De Roeck  \cite{NaDe04}  found in the Lorentz gauge. 
\item

We sketch some further developments of the scattering theory  associated with  a regular source. 
As    in (\ref{twist}) define $ \Pi_{\circ, \inn}( h )    =   \frac{i}{\sqrt{2}}  \B( a_{\inn}^*(h) - a_{\inn}(h)  \B ) $.
This is  essentially self-adjoint on $\cD$ and we define a unitary operator by 
\be
S = \exp \B(i \sqrt{2}  \Pi_{\circ, \inn} (J_T) \B)  =  \exp \B( -  a_{\inn}^*(J_T) + a_{\inn}(J_T) \B) 
\ee 
We have
\be
\begin{split} 
S^{-1} \Om_{\inn}  = &   \exp \B(     a^*_{\inn}(J_T) -  a_{\inn}(J_T)   \B )\Om_{\inn}   \\
 = & e^{- \frac12 \| J_T \|^2}   \exp\B (     a^*_{\inn}(J_T) \B)  \exp \B(    -  a_{\inn}(J_T)  \B)  \Om_{\inn}   \\
 = & e^{- \frac12 \| J_T \|^2}   \exp\B (      a^*_{\inn}( J_T) \B)  \Om_{\inn}   
 =  \Om_{\out} \\
\end{split}
\ee
and 
\be
S^{-1} a_{\inn}(h)  S =    a_{\inn}(h)   +  [ a_{\inn}^* (J_T)  , a_{\inn} (h) ]  =   a_{\inn}(h)   - (h,J_T)  = a_{\out} (h) 
\ee 
So the in and out fields are unitarily equivalent.

 One can form other asymptotic states  by applying creation operators   $a^*_{\out}  (h)$ to  $  \Om_{\out} $ and then  form scattering amplitudes
like 
\be
\B(  a^*_{\out}(h'_1) \cdots a^*_{\out} (h'_m ) \Om_{\out},  a^*_{\inn}  (h_1) \cdots a^*_{\inn } (h_n) \Om_{\inn} \B) 
\ee 
This can also be written 
\be
\B(  a^*_{\inn }(h'_1) \cdots a^*_{\inn } (h'_m ) \Om_{in },  S \ a^*_{\inn}  (h_1) \cdots a^*_{\inn } (h_n) \Om_{\inn} \B) 
\ee 
and $S$ is revealed as a scattering operator. 
\end{enumerate}

  \section{Point  source -  I }
  
  Again we study the equations $\sq A  =   j_T $  and $  \nabla \cdot A =0 $  and look for solutions which are free in the distant past
  and future. 
  We   now specialize to a  point source with current   $j(t,x) = x'(t) \de(x-x(t)) $.   We assume in this section that $x'(t) $ has compact support.
   So the particle is at rest in the distant past, accelerates for finite amount of time, and then stops in the distant future.  
      
  Again 
 we take $A_{\inn}  (f, t ) $ to be a free field as defined  in (\ref{ingot}) for real $f \in \cS(\bbR^3) $. 
 Then we define 
 \be    \label{zucchini2}
 A (f,t)   = A_{\inn} (f,t)  +  (  G_+  * j_T)( f,t) 
 \ee
Now we  do not  define   $ G_+  * j_T $ pointwise,   but  treat it as a distribution and    proceed by throwing
everything onto the test  function $f$. 
  Since 
 $G_+ (x,t) = G_-(-x,-t) $  we interpret  $(  G_+  * j_T)( f,t) $ as 
 \be \label{something1}  
    \blan G_+  * j_T,   f \otimes \de_t \bran 
  =  \blan  j_T,  G_- * ( f \otimes \de_t )  \bran 
   \ee 
 Then formally  we can transfer the transverse projection 
from  $j$ and to $f$  and get   
 \be \label{something2}  
    \blan  j_T,  G_- * ( f \otimes \de_t )  \bran 
  =  \blan  j,  G_- * ( f_T \otimes \de_t )  \bran 
   \ee 
 The latter is well-defined  for our current since    $ G_- * ( f_T \otimes \de_t ) $ as given by  (\ref{sister}) 
is a bounded function continuous in the spatial variable. 
   With this interpretation  (\ref{zucchini2}) becomes
  \be    \label{zucchini3}
 A (f,t)   = A_{\inn} (f,t)  +  \blan  j,  G_- * ( f_T \otimes \de_t )  \bran 
 \ee
Since  $j$ has compact support   and $ G_- * ( f_T \otimes \de_t )$ vanishes to the future of $t$, we have that $<  j,  G_- * ( f_T \otimes \de_t ) > $ vanishes
for $t$ sufficiently negative and so 
 $A (f,t)   = A_{\inn} (f,t)  $ in the distant past.    Since  $ A_{\inn} (f,t)  $ is free and since $G_{-}$ is a fundamental solution 
 we have 
 \be
 \begin{split} 
\sq A_{\out} (f,t) =    & \frac{ \pa^2 A_{\out} }{ \pa t^2} (f,t) - A_{\out} ( \De f, t)  \\
  =& \blan j,  \frac{\pa^2}{ \pa t^2} G_- * ( f_T \otimes \de_t )  -   G_- * ( ( \De f)_T \otimes \de_t ) \bran \\
   = &  \blan j, \sq  (  G_- * ( f_T \otimes \de_t ) )   \bran=  \blan j,   f_T \otimes \de_t  \bran  \equiv
     \blan j_T,   f \otimes \de_t  \bran \\
 \end{split}
  \ee
So $A$  satisfies $\sq A  =   j_T $ in the sense of distributions.   Similarly $\nabla \cdot A =0$.

  Next   we define $ A_{\out }  (f,t)  $ by     
 \be \label{squash2} 
 A (f,t)   = A_{\out } (f,t)  +   G_- *  j _T(f,t) 
  \ee
which   is interpreted as
   \be \label{squash3} 
 A (f,t)   = A_{\out } (f,t)  +   \blan  j,  G_+ * ( f_T \otimes \de_t )  \bran 
  \ee
 Then   $ A (f,t)   = A_{\out} (f,t) $ in the distant future.

Combining (\ref{zucchini3}) and (\ref{squash3}) gives 
  \be
  A_{\out } (f,t)  =   A_{\inn }  (f,t)  -   \blan  j,  G* ( f_T \otimes \de_t )  \bran 
 \ee
 Then      $ A_{\out }$ satisfies      $ \sq A_{\out}   =  0 $ and   $\nabla \cdot A_{\out}  =0$ in the sense of distributions 
 and is a free field.

 Now take the expression  $( G *  (f_T \otimes \de_{t}) )(x,s ) =  u_{f_T} (x,s-t) $ from (\ref{sharp})
 and  find 
 \be \label{nuovo} 
 \begin{split} 
  \blan  j,  G* ( f_T \otimes \de_t )  \bran 
  =  &   \int x'(s) \de(x-x(s)) u_{f_T} (x,s-t) dx ds  \\
  =   &   \int x'(s)   u_{f_T} (x(s) ,s-t)  ds \\
    = &    (2 \pi )^{-\frac32}   i  \int  x'(s)   \B(  e^{-i \om (s- t )}   - e^{i \om (s-t) }   \B)e^{ipx(s) }   \tilde f_T(p)     \frac{dp}{ 2 \om} \ ds \\
       = &    (2 \pi )^{-\frac32}    \int  x'(s)   \B[ i e^{-i \om (s- t )+ipx(s) }\tilde f_T(p)       - ie^{i \om (s-t) -ipx(s) }   \tilde f_T(-p)   \B]   \frac{dp}{ 2 \om} \ ds \\
      = &    (2 \pi )^{-\frac32}  2 \ \re \   \int  x'(s)   \B[- ie^{i \om (s-t) -ipx(s) }   \tilde f_T(-p)  \B]  \  \frac{dp}{ 2 \om}   ds \\
  \end{split} 
 \ee
Now change the order of integration (easily justified),   and define  (new definition) 
 \be \label{deep} 
 J(p)  = -i  (2 \pi )^{-\frac32}    \int  x'(s)     e^{i \om s - ipx(s) }  ds 
\ee 
 Then  
 \be \label{pixie} 
   \blan  j,  G* ( f_T \otimes \de_t )  \bran  =  2 \ \re \   \int     e^{-i \om t}  \tilde f_T(-p)   J(p)    \frac{dp}{ 2 \om}   =2 \ \re \  (e^{i\om t} \tilde f _T, J_T)  
 \ee
  Then integral converges since $J_T(p)$ is bounded and $\tilde f_T(p)$ is bounded and rapidly decreasing.   We have written
  it as an inner product in $\cH$ but are not  asserting that $J_T \in \cH_T$, at least not yet.  
  Now we have as before
  \be \label{oranges}
  A_{\out } (f,t)  =   A_{\inn }  (f,t)  -  2 \ \re \  (J_T,e^{i\om t} \tilde f _T)  
 \ee

  We consider   the  linear function $h \to (J_T,h)$   for  
   $h$ in a dense  domain $\cS \subset \cH_T$  containing  the functions $e^{i \om t} \tilde f_T$ for $f \in\cS(\bbR^3)$. 
  It is the domain of of rapidly decreasing functions: 
   \be 
 \cS = \{ h \in \cH_T:  \textrm{ for  } k \geq 0 \textrm{ there is }  C \textrm{  so } |h((p)| \leq C(1+ |p|^2)^{-k}   \}
 \ee
For  $h \in \cS$ we   define 
    \be  \label{xmas4}
   \begin{split} 
 a^*_{\out} (h)  = &  a^*_{\inn} (h) - (J_T,h)   \\  
 a_{\out} (h)  =&   a_{\inn} (h)  -  (h, J_T)   \\ 
 \end{split} 
 \ee
 and then as in (\ref{xmas1}) - (\ref{xmas3})
 \be  \label{xmas5}
  A_{\out} (f,t)  = a^*_{\out} (e^{i \om t} \tilde f_T)+ a_{\out} (e^{i \om t} \tilde f_T)
\ee 
just  as for     $ A_{\inn } (f,t)  $.

Now   we can write $A_{\inn/\out} (f,t)  = \sqrt{2} A_{\circ,\inn/\out}(e^{i \om t} \tilde f_T  ) $ where the momentum space fields are  
  $A_{\circ,\inn/\out}( h)   =  \frac{1}{\sqrt{2}}  ( a_{\inn/\out}^*(h) + a_{\inn/ \out }(h) )$.  In terms of these fields 
 the defining relation (\ref{xmas4})  can be written
 \be
  A_{\circ, \out}(h) = A_{\circ,\inn}( h) - \sqrt 2  \  \re \  (J_T, h)  
\ee
 With this identity  we pass to the $C^*$ algebra.  Define
\be
   W_{ \out } (h)  =  e^{iA_{\circ,\out}( h)}  \hs
W_{  \inn } (h)  =  e^{iA_{\circ,\inn}( h)}   
  \ee
and then 
\be
W_{ \out  } (h) = W_{ \inn} (h) e^{- i \sqrt 2  \  \re \  (J_T, h)  }
\ee
These are elements of  $C^*$ algebra  $\EuFrak{A} $ generated by  $ W_{ \inn} (h)$,   $h \in \cS$.  
 In the Fock vacuum $\om_{in} = ( \Om_{0, in} , [ \ \cdot. \ ]   \Om_{0, in}) $ we have
\be \label{honey} 
\om_{in} ( W_{\out} (h)   )  = \om_{\inn} ( W_{\inn} (h)  )   e^{-i \sqrt{2}  \ \re  ( \bar J_T, h ) }
\ee
 
But $W_{\out} (h) $ is again a representation of the CCR and generates the same $C^*$ algebra $\EuFrak{A} $ as $W_{\inn }(h) $.
Then there is a unique  $*$-automorphism $\al$ on $\EuFrak{A} $  such that $\al ( W_{\out} (h) )  = W_{\inn } (h)$ (see for example \cite{BrRo81}).  
This satisfies $\al ( W_{\inn} (h) )  = W_{\inn } (h)  e^{i \sqrt{2}  \ \re  ( \bar J_T, h ) }$. 
 If we define a
new  state  by   $\om_{\out} \equiv  \om_{\inn}  \circ \al $    then 
\be \label{honey2} 
\om_{\out} ( W_{ \inn} (h)   )  = \om_{\inn} ( W_{ \inn } (h)  )    e^{i \sqrt{2}  \ \re  ( J_T, h ) }
\ee
Instead of changing operators we have changed states.  The  form of the equation shows that $\om_{\out}$ is a coherent state by 
our $\cC^*$ algebra definition.

The question is now whether $\om_{\out} $ is a Fock state.
Before answering the question we sharpen our criterion for when this is the case.

\begin{prop} 
$\om_{\out} $ is a Fock state if and only if  $J_T  \in \cH_T$. 
\end{prop} 
\bigskip

\pr  We have already noted that  $\om_{\out} $ is a Fock state if and only if the linear function
$h \to (J_T,h)$ is continuous on the pre-Hilbert space $\cS$.  We now claim this is equivalent to $J_T \in \cH_T$.
If $J_T \in \cH_T$ then continuity is immediate.    If the continuity holds then there is $J' \in \cH_T$ such that  
$(J_T,h)= (J',h)$ for $h \in \cS$.   If $h$ rapidly decreasing then $h_T  \in \cS$ and  $(J_T- J', h) =(J_T- J', h_T)=0 $. Then 
 $(J_T- J', h) =0 $  for $h \in \cS(\bbR^3)$.  Hence  $(J_T - J')/2 \om$ as distributions and hence almost everywhere as functions.   
Thus $J_T = J'$ almost everywhere  and so $J_T \in \cH_T$.  This completes the proof. 
\bigskip 
 
 In the following  we  assume that the velocity of the charged  
 particle is less than the speed of light which is here one.

\begin{prop}  \label{inca}  Let $x(t) $  be 
$ \cC^3  $   with  $|x'(t)| \leq v_0<1$, and suppose $x'(t) $ has compact support.  
Then   $ J_T \in \cH_T $ and $\om_{\out}$ is the Fock state  defined by $\Om_{\out} = \Om_{J_T}$. 
\end{prop} 

\pr    It suffices to show  $J \in \cH$.    Ignoring the multiplicative constants  in  (\ref{deep}) we might as well assume that
\be  \label{54} 
J   (p)  =      \int    x '(t)  e^{i  \psi(p,t)   }   dt   \hs \hs \psi(p,t)  =   \om t - p \cdot x(t)  
\ee
Since  this is a bounded continuous function of $p$,   
to decide whether it is in $\cH $   we need to study asymptotics as $p \to \infty$.  Our first thought might be the stationary phase method.  So 
we look for points where  
\be
\psi_t( p,t)  =  \om - p \cdot x'(t)  =0
\ee 
   However $ |p \cdot x'(t) | < |p|v_0= \om v_0 $ and   
\be
 \om ( 1-v_0)  \leq   |\psi_t( p,t)|  \leq  \om ( 1+v_0) 
\ee
There is no point of stationary phase  for $p \neq 0 $.   
Instead we  can integrate by parts and write with $\psi_{tt} = - p \cdot x''(t) $
\be \label{parts1}
\begin{split}
J  (p)  =   &     \int  x'(t)   \frac{ 1}{ i \psi_t}  \frac{\pa}{ \pa t}  e^ {i  \psi(p,t)   }  dt \\
=&    \frac{1}{i} \int  \left[  \frac{\psi_{tt} }{  \psi^2_t}  x'(t)  -  \frac{ 1}{  \psi_t}  x''(t)  \right]e^ { i  \psi(p,t)   }  dt \\
\end{split} 
\ee
Now $x', x''$ are bounded and   $\psi_t^{-1} $  and  $\psi_{tt} /  \psi^2_t $ are $\cO(\om^{-1} )$ for $|p| \geq 1$.  Hence $J  (p) =   \cO(\om^{-1} )$.   This is not quite enough for convergence,
but we can repeat the last step and get 
\be \label{parts2}
\begin{split}
J   (p)  =   &
  \frac{1}{i} \int  \left[  \frac{\psi_{tt} }{  \psi^2_t}  x'(t)  -  \frac{ 1}{  \psi_t}  x''(t)  \right] \frac{ 1}{ i \psi_t}  \frac{\pa}{ \pa t}  e^ {i  \psi(p,t)   }  dt \\
=  &  \int  \left[ \B(  \frac{\psi_{ttt} }{  \psi^3_t} - \frac{3\psi_{tt} ^2 } {\psi_t^4} \B) x'(t)  +  \frac{ 3\psi_{tt} }{  \psi^3_t}   x''(t) 
- \frac{ 1}{\psi_t^2} x'''(t)  \right]e^ { i  \psi(p,t)   }  dt 
\end{split}
\ee
Now   $x', x'', x'''$ are all bounded and the coefficients  are all $\cO(\om^{-2})  $ for $|p| \geq 1$.    Thus $J(p) = \cO( \om^{-2} )$ and  we have
\be
\int _{|p| \geq 1}   | J   (p)  |^2  \frac{dp} {2 \om } \leq  \const   \int _{|p| \geq 1} \om^{-5} dp < \infty
\ee 
Hence $J \in \cH$ and the proof is complete. 
\bigskip

\rem  
We can allow jump discontinuities in $x'''(t)$ without changing the result. 
 Suppose we allow a jump discontinuity in  $x''(t)$, say at   $t=0$.
Then in the integration by parts in (\ref{parts2})   we get a boundary term which is 
\be
\left[ \frac{ p \cdot (x''(0^-) - x''(0^+) )x'(0) }{ \psi^3_t(p,0)}  +   \frac{    x''(0^- ) - x''(0^+ ) }{  \psi^2_t(p,0)} \right]   e^ {i  \psi(p,0)}
\ee
This is still $\cO(\om^{-2} ) $ and our results still hold.    However  a discontinuity in the velocity $x'(t)$ 
is more serious as we now show. 
 
\begin{prop}  \label{stinger2}
Suppose $x(t)$ is continuous on $\bbR$   and  $\cC^3$ on $(- \infty ,0] $ and $[0,\infty) $ with  
\be
   x'(0^-) - x'( 0^+ ) \neq 0
 \ee 
Suppose   also   $x'(t) $ has compact support and   $|x'(t)| \leq v_0<1$.  
Then  $J_T  \notin \cH_T $ and $\om_{out} $ is not 
 a Fock state
\end{prop} 
\bigskip

\pr  Since $J(p)$ is still bounded, it  again suffices to consider asymptotics as $p \to \infty$.   In the integration by parts in (\ref{parts1}) there is now  a boundary term and we have
\be \label{parts3}
\begin{split}
J (p)  =   &     \int  x'(t)   \frac{ 1}{ i \psi_t}  \frac{\pa}{ \pa t}  e^ {i  \psi(p,t)   }  dt \\
=&    \frac{1}{i} \int  \left[  \frac{\psi_{tt} }{  \psi^2_t}  x'(t)  -  \frac{ 1}{  \psi_t}  x''(t)  \right]e^ { i  \psi(p,t)   }  dt   + \frac{1}{i}  \De (p) e^ {i  \psi(p,0)}  \\
\end{split} 
\ee
where
\be
\De  (p)  =     \frac{ x'(0^- ) }{  \om - p\cdot x'(0^- )}  -    \frac{ x'(0^+  ) }{  \om - p\cdot x'(0 ^+  )}
\ee
The first term in (\ref{parts3})  is analyzed as in the previous proposition,  unaffected by possible discontinuities in $x'', x'''$, and the transverse part is
in $\cH_T$.   Thus to prove our result it suffices to show that $\De_T$ is not in $\cH_T$.

Let $v^{-} = x'(0^- )$ and $v^+ = x'(0^+) $. Then this can be written
\be
\De(p) = \frac{  \om (v^- - v^+ )    + ( v^+ p \cdot v^- -   v^- p \cdot v^+ )  } { ( \om - p\cdot v^-) ( \om - p\cdot v^+)  }
\ee
Now $ v^+ p \cdot v^- -   v^- p \cdot v^+  =   p  \times  ( v^+ \times  v^- ) $. If we define $\de v = v^--v^+ \neq 0 $ and 
 $n_p = p/\om = p /|p| $  then  $\De  (p) = \De_1(p)  + \De_2(p) $ where
\be
\begin{split} 
\De_1  (p)  =   &  \frac{1}{\om}  \left[ \frac{ \de v     } { ( 1 -n_p\cdot v^-) (1 - n_p\cdot v^+)  }\right] \\
\De_2  (p)  =   &  \frac{1}{\om}  \left[ \frac{   n_ p  \times  ( v^+ \times  v^- )   } { ( 1 -n_p\cdot v^-) (1 - n_p\cdot v^+)  }\right] \\
\end{split} 
\ee
Note that the denominators here are bounded below since $|v^{\pm} | <1$. 
Now $\De_2(p) $ is already transverse,  and the transverse part of $\De_1(p) $ is with $n_{\de v} = \de v/|\de v| $ 
\be 
 \De_{1,T}   (p)  =     \frac{1}{\om}  \left[ \frac{ \de  v - n_p (n_p\cdot \de v )     } { ( 1 -n_p\cdot v^-) (1 - n_p\cdot v^+)  }\right]
=  \frac{ |\de v|  }{\om}  \left[ \frac{n_{ \de  v }  - n_p (n_p\cdot n_{\de v } )   } { ( 1 -n_p\cdot v^-) (1 - n_p\cdot v^+)  }\right]
\ee
We have now  $\De_T  (p) = \De_{1,T} (p)  + \De_2(p)$.

Now we divide into two cases.  Either $v^{\pm} $ are colinear or not.  If they are colinear  then 
$\De_2 (p) =0$, and we compute
\be \label{pasta1}
\begin{split} 
\int_{|p| \geq 1}  |\De_T  (p) |^2 \frac{dp}{2\om}  
&\geq    \int_{|p| \geq 1, |n_p\cdot n_{\de v } |  \leq \frac12 }    |\De_{1,T}   (p) |^2 \frac{dp}{2\om}  \\ 
&\geq \const  |\de v|^2   \int_{|p| \geq 1,| n_p\cdot n_{\de v }|
   \leq \frac12 }   \frac{1 }{\om^3 } dp = \infty  \\ 
\end{split}
\ee

 If $v^{\pm}$ are not colinear then  $v_{\times} \equiv  v^+ \times  v^-  \neq 0$.   We look at the component of  $\De_T(p)$ along $v_{\times}$.
 Again $\De_2(p) $ does not contribute, nor does the $n_{\de v}$ term in $\De_{1,T} (p)$.  We have
 \be \label{outback} 
n_{ v_{\times } } \cdot   \De_T  (p)  =   \frac{ |\de v|  }{\om}  \left[ \frac{ -  ( n_p\cdot  n_{v_{\times}}) (n_p\cdot n_{\de v } )   } { ( 1 -n_p\cdot v^-) (1 - n_p\cdot v^+)  }\right]
\ee
Then 
\be \label{pasta2} 
\begin{split} 
\int_{|p| \geq 1}  |\De_{T } (p) |^2 \frac{dp}{2\om}  
 & \geq     \int_{|p| \geq 1,    |n_p \cdot  n_{ v_{\times} }| \geq \frac12,  | n_p   \cdot  n_{\de v}|  \geq \frac12     } |n_{ v_{\times} }  \cdot \De_{T } (p) |^2 \frac{dp}{2\om}   \\
& \geq \const  |\de v|^2  \int_{|p| \geq 1,    |n_p \cdot  n_{ v_{\times} }| \geq \frac12,  | n_p   \cdot n_{ \de v}|  \geq \frac12     } \frac{1 }{\om^3 } \ dp = \infty  \\ 
\end{split}
\ee
The last step follows since $\de v$ orthogonal to   $v_{\times}$ implies   the cones   $| n_{v_{\times}}  \cdot  n_p | \geq \frac12$  and  $|n_p \cdot n_{ \de v} | \geq \frac12 $ have a non-empty intersection.  This completes the proof.

 \section{Point source - II  }

  Now we consider a trajectory $x(t)$  which has constant velocity $v_{\inn} $ in the distant past, and constant velocity $v_{\out} $ in the distant future, and
is otherwise smooth.      So we no longer have compact support for $x'(t)$, but we do have compact support for $x''(t)$. 

There is now a problem with the convergence of integrals like (\ref{deep}).   To avoid this we approximate the current $j$  by
\be
j_R (x,t) = \begin{cases} x'(t) \de ( x- x(t) )  & \hs |t| \leq R \\ 
 0 & \hs \textrm{ otherwise } \\
 \end{cases} 
 \ee
and seek to take the limit $R \to \infty$.   
The analysis proceeds as in the the  previous section and we have $ A_{\out } (f,t)  =   A_{in }  (f,t)  -   \blan  j_R,  G* ( f_T \otimes \de_t )\bran $
where   
 \be \label{nuovo2} 
 \begin{split} 
  \blan  j_R,  G* ( f_T \otimes \de_t )  \bran 
        = &    (2 \pi )^{-\frac32}  2 \ \re \   \int_{-R}^R  x'(s)  \int \B[- ie^{i \om (s-t) -ipx(s) }   \tilde f_T(-p)   \B]   \frac{dp}{ 2 \om} \ ds \\
  \end{split} 
 \ee

\begin{prop} Let $x(t) $  be 
$ \cC^3  $ with  $|x'(t)| \leq v_0<1$, and  suppose $x''(t) $ has compact support. 
Then the limit $R \to \infty$ exists  on the right side of   (\ref{nuovo2}) and defines the left side at $R= \infty$.  We have as in (\ref{pixie})  
\be \label{lamb1}
  \blan  j,  G* ( f_T \otimes \de_t )  \bran    = 2 \  \mathrm{Re}  \     \int  e^{-i \om t}  \tilde f_T(-p)  J (p)  \frac{dp}{2 \om} =  2 \  \mathrm{Re}  \   (e^{i\om t}\tilde f_T, J_T)  
\ee
where up to a multiplicative constant  
\be \label{lamb2}
J (p)
=     \frac{1}{i} \int  \left[  \frac{\psi_{tt} }{  \psi^2_t}  x'(t)  -  \frac{ 1}{  \psi_t}  x''(t)  \right]e^ { i  \psi(p,t)   }  dt \ee
 as in (\ref{parts1}),    and the representation (\ref{parts2}) for $J(p)$  holds as well. 
\end{prop}
\bigskip

\pr  Note that since $\psi_{tt} = - p \cdot x''(t)$   the integrand on the right side of  (\ref{lamb2})  has compact support in $t$ and  that integral exists.  
Furthermore $J(p) $ is $\cO(\om^{-1} ) $  everywhere   and   $\tilde f_T(p) $ is bounded and rapidly decreasing so the integral  (\ref{lamb1}) converges.   Once 
(\ref{lamb2}) is established (\ref{parts2}) holds
by a further integration by parts.

To derive (\ref{lamb1})  it suffices to consider the case  $t=0$.  Then   in (\ref{nuovo2})    with $\psi(p,s)  = \om s - p \cdot  x(s)$ we need to find an   $R \to \infty$ limit for 
\be \label{ocho}
\begin{split}
&    \int _{-R}^R   x'(s) \left[ \int  e^ {i  \psi(p,s)   } \tilde f_T(-p)   \frac{dp}{2\om}  \right]       ds \\
 =   &     \int _{-R}^R   x'(s)  \frac{d}{ d s} \left[ \int  \frac{ 1}{ i \psi_s}   e^ {i  \psi(p,s)   }  \tilde f_T(-p)   \frac{dp}{2\om} \right]       ds 
+    \int _{-R}^R   x'(s)   \left[ \int   \frac{ \psi_{ss} }{ i \psi^2_s}   e^ {i  \psi( p,s)   } \tilde f_T(-p)  \frac{dp}{2\om}  \right]       ds \\   
= &   \frac{1}{i} \int_{-R}^R \left[  \int  \left(  \frac{\psi_{ss} }{  \psi^2_s}  x'(s)  -  \frac{ 1}{  \psi_s}  x''(s)  \right)e^ { i  \psi(p,s)   } \tilde f_T(-p)   \frac{dp}{2\om}  \right]   ds  \\
& +       \left[  x'(s) \int  \frac{ 1}{ i \psi_s}   e^ {i  \psi(p,s)   } \tilde f_T(-p)   \frac{dp}{2\om}  \right]_{s= -R}^{s=R}                       \\
\end{split} 
\ee
In the double integral the $R \to \infty$ limit exists by the support of $x''(t)$.  The integrand is $\cO(\om^{-2})$ as  $|p| \to 0$ and rapidly decreasing as $|p| \to \infty$, so 
the integral is  is absolutely convergent.  Changing the order of integration gives  the result (\ref{lamb1}), provided the endpoint term goes to zero.

For the endpoint term we need to show     
 $ \int    \psi_t^{-1}  e^{i\psi(t,p) }  \tilde f_T(-p)   dp/2\om$ 
 goes to zero as $|t|  \to  \infty$.   As $t  \to  \infty $ we have  $x'(t) \to v_{\out} $ so it suffices to
 show    
\be   \label{loony} 
  \int  \frac{1} {  \om  - p\cdot v_{\out} }    e^{i ( \om - p\cdot v_{\out}  ) t }  \tilde f_T(-p)  \frac{ dp }{2 \om} 
  \ee
  goes to zero as $t  \to \infty$.   Let $p \to - p$ and then
  write this in polar coordinates $(|p|, \theta, \phi)$ with $v_{\out} $ as the $z$-axis.  Then   $ \om + p\cdot v_{\out}  = \om ( 1+\cos \theta \ v_{\out} ) $
  and   $ ( 1+  \cos \theta \ v_{\out} ) $ is bounded above and below.   The integral is now
\be   \label{loony2} 
  \int    \frac{1} {  2 ( 1+ \cos \theta\  v_{\out} )  } \B[ \int   e^{i |p| ( 1+  \cos \theta \ v_{\out} )  t }  \tilde f_T(|p|, \theta, \phi ) \  d|p| \B]\sin \theta d \theta d \phi  
  \ee
 Now the interior integral goes to zero by the Riemann- Lebesgue lemma,  and the overall integral goes to zero by dominated convergence.  The limit $t \to - \infty$
 is entirely similar. 
 This completes the proof. 
 \bigskip
 
 Having established (\ref{lamb1}), (\ref{lamb2}) we proceed as in (\ref{oranges}) - (\ref{honey2})     and again find
\be
 \om_{\out} ( W_{ \inn} (h)   )  = \om_{\inn} ( W_{ \inn } (h)  )    e^{i \sqrt{2}  \ \re  ( J_T, h ) } 
  \ee

\begin{prop}  \label{jolt} 
  Let  $x(t)$ be $\cC^3$  with  $|x'(t)| \leq v_0<1$, and  suppose  $x'(t) = v_{\inn}$ in the distant past and $x'(t) =v_{\out} $ in the distant future so $x''(t)$ has compact support.   
Then  $\om_{\out}$ is a Fock state if and only if  $v_{in}  =  v_{\out}$.
\end{prop}
\bigskip

\pr
We must show that $J_T \in \cH_T$ if and only if $v_{in}  =  v_{\out}$. 
We can use the representation (\ref{parts2})  as before  
to  get convergence for $|p| \geq 1$.   So the issue is whether the  
the integral for $|p| \leq 1$ converges.  (In proposition  \ref{inca}  convergence  for $|p| \leq 1$ was easy.)     Our general bound is that 
 $J( p)  $ is $\cO(\om^{-1})$, as in $J_T(p)$. 
We show that  in case  $v_{in} \neq v_{\out} $ this is actually sharp and gives a logarithmic divergence. 

In the representation (\ref{lamb2})   for $|p| \leq 1$ we write
$e^{i \psi(p,t) } = 1 + \cO(\om) $ and correspondingly   $J(p) = F (p) + \cO(1)$.  The $\cO(1) $ term  is in $\cH$ and the transverse part in in $\cH_T$,
so it suffices to consider $F(p)  $ and show that $F_T(p)  $ either is or  is not  square integrable for $|p| \leq 1$ with respect to $(2\om)^{-1} dp$. 

We have up to a multiplicative constant
\be
\begin{split} 
F (p)
=   &   \int  \left[  \frac{x'(  p \cdot x'' )  }{   \om - p \cdot x'(t)  ) ^2}   +  \frac{ x''}{ (   \om - p \cdot x'(t))  }     \right]   dt   \\ 
= &    \int  \frac{d}{d t }    \left[
\frac{x'(t) } {( \om - p \cdot x'(t))  }  \right]  dt  \\
= &     
\frac{v_{\out} } {(  \om - p \cdot v_{\out} )  }  -   \frac{v_{\inn}  } { ( \om - p \cdot v_{\inn} )  } 
\\
= &     
\frac{ \om  ( v_{\out} - v_{\inn} )  +(v_{\inn}\  p\cdot v_{\out}-  v_{\out}\  p\cdot v_{\inn }  ) } 
{ ( \om - p \cdot v_{\out}  ) (  \om - p \cdot v_{\inn} )  } \\
\end{split}
\end{equation}
where  $\de  v= v_{\out} - v_{\inn} $.
We see that if $v_{\inn} = v_{\out} $ then $F  = 0$ and $F_T =0$  and hence $J \in \cH$ and  $J_{T} \in \cH_T$ and $\om_{\out}$ is Fock.

We proceed  with the case  $  v_{\out} \neq   v_{\inn } $.    We are now almost exactly in the situation that occurred in the proof of   Proposition \ref{stinger2}. The 
main difference is that we consider $|p|  \leq 1$ rather than $|p| \geq 1$.     We  write  $F_T(p) = F_{1,T} (p) + F_2(p)  $  
where with $\de v \equiv  v_{\out} -  v_{\inn } $ and   $  v_{\times} =  v_{\inn} \times v_{\out}   $
\be
 \begin{split}
 F_{1,T}   (p) = &   \frac{ |\de v|  }{\om}  \left[ \frac{n_{ \de  v }  - n_p (n_p\cdot n_{\de v } )   } {  (1 - n_p\cdot v_{\out} ) ( 1 -n_p\cdot v_{\inn} ) }\right] \\
F_2  (p)  =   &  \frac{1}{\om}  \left[ \frac{   n_ p  \times  n_{ v_{\times}  }   } { (1 - n_p\cdot v_{\out}) ( 1 -n_p\cdot v_{\inn} ) )  }\right] \\
\end{split}
\ee
If    $v_{\inn} $ and $v_{\out} $ are colinear  (which  includes the case where one of them is zero), then only $F_{1,T}$ contributes and we have
as in (\ref{pasta1})
\be \label{pasta1a}
\int_{|p| \leq  1}  |F_T  (p) |^2 \frac{dp}{2\om}   \geq \const  |\de v|^2   \int_{|p| \leq 1, |n_p\cdot n_{\de v }|
   \leq \frac12 }   \frac{1 }{\om^3 } dp = \infty  
\ee
If    $v_{\inn} $ and $v_{\out} $ are not  colinear we look at the component along $v_{\times}$ and find as in (\ref{outback})
 \be
n_{ v_{\times } } \cdot   F_T  (p)  =  
 \frac{ |\de v|  }{\om}  \left[ \frac{ -  ( n_p\cdot  n_{v_{\times}}) (n_p\cdot n_{\de v } )   } { (1 - n_p\cdot v_{\out} )  ( 1 -n_p\cdot v_{\inn} ) }\right]
\ee
Then as in (\ref{pasta2})
\be \label{pasta2a} 
\begin{split} 
\int_{|p| \leq 1}  |F_{T } (p) |^2 \frac{dp}{2\om}  
& \geq \const  |\de v|^2  \int_{|p| \leq 1,    |n_p \cdot  n_{ v_{\times} }| \geq \frac12,  | n_p   \cdot n_{ \de v}|  \geq \frac12     } \frac{1 }{\om^3 } \ dp = \infty  \\ 
\end{split}
\ee
Thus the $F_T$ condition fails, and $\om_{\out}$ is not  Fock.  
\bigskip

 \rems
 \begin{enumerate}

 \item  Note that if  the particle has constant velocity  then $x''(t) =0$, hence $J(p) = 0$,  and $\om_{\out} = \om_{\inn} $.   No photons
 are radiated.

 \item  The result does not depend on the complete history of the trajectory,  only on the initial and final states.
 Except in a the special circumstance $v_{\inn}  = v_{\out}$ we always leave Fock space.

 \item It is noteworthy that the departure from Fock space  comes from a UV divergence in the case of  of discontinuities in the velocity
 (proposition \ref{stinger2}), and from an IR divergence in the case $v_{\inn} \neq v_{\out} $(proposition \ref{jolt}).

 \item 
 It might be of interest to obtain results of this kind  for the radiation of gravitons by a classical source  in linearized quantum gravity. 
 See  Skagerstam, Eriksson, Rekdal  \cite{SER19b}, for a formulation of this model in an analog of the Coulomb gauge.      In this connection we  also mention the work of Kegeles, Oriti, Tomlin  \cite{KOT18} who suggest non-Fock coherent states
 of the type used here as appropriate for a group field theory model of quantum gravity. 
 \end{enumerate}

\end{document}